\documentclass[twocolumn]{aastex63}

\usepackage{graphicx}% Include figure files
\usepackage{dcolumn}% Align table columns on decimal point
\usepackage{bm}% bold math
\usepackage{ulem}

\begin{document}

\shorttitle{New Limits on Axionic Dark Matter}
\shortauthors{Darling}

\title{New Limits on Axionic Dark Matter from the Magnetar PSR J1745$-$2900}

\author{Jeremy Darling}
 \affiliation{Center for Astrophysics and Space Astronomy \\
Department of Astrophysical and Planetary Sciences \\
University of Colorado, 389 UCB \\
Boulder, CO 80309-0389, USA}
 \email{jeremy.darling@colorado.edu}


\begin{abstract}
  Axions are a promising dark matter candidate that were motivated to solve the strong {\it CP} problem and that may also address the
  cosmological matter-antimatter asymmetry.   Axion-photon conversion is possible in the presence of the strong magnetic fields, and the
  photon so produced will have energy equal to the axion mass.  Here we report new limits on axionic dark matter obtained from radio
  spectra of the Galactic Center magnetar PSR J1745$-$2900.  The magnetar has a magnetic field of $1.6\times10^{14}$~G that interacts with
  a dark matter density $2\times10^5$ to $2\times10^9$ times greater than the local dark matter encountered by terrestrial haloscopes,
  depending on the Galactic dark matter profile.
  No significant spectral features are detected across 62\% of the axion mass range 4.1--165.6~$\mu$eV (1--40~GHz).  The interpretation of flux
  limits into limits on the two-photon coupling strength $g_{a\gamma\gamma}$ depends on the magnetospheric conversion model and
  on the dark matter density at the Galactic Center. For a standard dark matter profile, we exclude axion models with
  $g_{a\gamma\gamma}> $ 6--34~$\times 10^{-12}$~GeV$^{-1}$ with 95\%
  confidence over the mass ranges 4.2--8.4, 8.9--10.0, 12.3--16.4, 18.6--26.9, 33.0--62.1, 70.1--74.3, 78.1--80.7, 105.5--109.6, 111.6--115.2,
  126.0--159.3, and 162.5--165.6~$\mu$eV.
  For the maximal dark matter cusp allowed by stellar orbits near Sgr A*, these limits reduce to $g_{a\gamma\gamma} > $ 6--34 $ \times10^{-14}$~GeV$^{-1}$, 
  which exclude some theoretical models for masses $> 33$~$\mu$eV.
  Limits may be improved by modeling stimulated axion conversion, by ray-tracing
  conversion pathways in the magnetar magnetosphere, and by obtaining deeper broad-band observations of the magnetar.
\end{abstract}

%\keywords{astroparticle physics --- dark matter --- elementary particles --- pulsars: general --- pulsars: individual (PSR J1745$-$2900) --- stars: magnetars  --- Galaxy: center}

\section{\label{sec:intro}Introduction}

The Peccei-Quinn mechanism offers a solution to the strong $CP$ (charge-parity) problem in quantum chromodynamics (QCD) with the introduction of the axion,
a spin zero chargeless massive particle \citep{Peccei1977,weinberg1978,wilczek1978}.  Moreover, QCD axions are a promising cold dark matter candidate \citep{preskill1983,dine1983,abbott1983}, and may explain the matter-antimatter asymmetry in the early universe \citep{Co2020}.

If axions ($a$) exist, they would have a two-photon coupling $g_{a\gamma\gamma}$ such that the electromagnetic interaction is
\begin{equation}
\mathcal{L}_{a\gamma\gamma} = -(1/4) g_{a\gamma\gamma} a\, F_{\mu\nu}\tilde{F}^{\mu\nu} =  g_{a\gamma\gamma} a\, {\mathbf E}\cdot{\mathbf B}.
\end{equation}
Axion-photon conversion can thus occur in the 
presence of a magnetic field, but the axion-photon coupling is weak: $g_{a\gamma\gamma}\sim 10^{-16}$~GeV$^{-1}$
for axion mass $m_a = 1$~$\mu$eV \citep{Kim1979,Shifman1980,Dine1981,Zhitnitsky1980,sikivie1983}.
Theory predicts $g_{a\gamma\gamma} \propto m_a$, 
but the mass is not constrained.  Axion searches must therefore span decades in $m_a$ while reaching very small $g_{a\gamma\gamma}$.

Axion experiments include CAST, which searched for Solar axions \citep{arik2014,arik2015}, and ``haloscopes'' 
that use narrow-band resonant cavities to detect dark matter axions, such as ADMX and HAYSTAC
\citep{asztalos2001,asztalos2010,brubaker2017,Zhong2018}.
There are also natural settings where one may use telescopes to conduct sensitive and wide-band QCD axion searches
toward pulsars or galaxies \citep{hook2018,huang2018,day2019,leroy2020,battye2020,edwards2020,mukherjee2020}.

Of particular interest is the natural axion-photon conversion engine, the Galactic Center magnetar PSR J1745$-$2900.
PSR J1745$-$2900 has a strong magnetic field \citep[$1.6\times10^{14}$~G;][]{mori2013},
and it should encounter the highest dark matter flux in the Galaxy \citep{hook2018}.
Axions will encounter a plasma frequency at some radius in the magnetosphere that equals
its mass, allowing the axion to resonantly convert into a photon \citep{hook2018}.
The likely axion mass range, 1--100~$\mu$eV, equates to radio frequencies 240 MHz to 24 GHz.

In this {\it Letter}, we present archival observations of PSR~J1745$-$2900 from the NSF's
Karl G. Jansky Very Large Array (VLA\footnote{The National Radio Astronomy Observatory is a facility of the National Science Foundation operated under cooperative agreement by Associated Universities, Inc.}) that expand upon our previous work \citep{darling2020}. 
We obtain 95\% confidence limits on resonant axion-photon conversion emission line flux density from the magnetar
spanning 62\% of the 1--40~GHz band.  Limits on the axion-photon coupling, $g_{a\gamma\gamma}$, rely on a neutron star
magnetosphere model and are bracketed by two limiting-case uncored Galactic dark matter profiles.
We present model caveats, discuss observational limitations, and suggest 
observational and theoretical work to expand the $g_{a\gamma\gamma}$ vs.\ $m_a$ space probed by this technique.

\begin{deluxetable*}{ccccccccrcl}
  \tablecaption{\label{tab:obs}Very Large Array Summary of Observations}
%\begin{ruledtabular}
%\begin{tabular}{ccccccccrcl}
  \tablehead{  \colhead{Band} & \colhead{Frequency} & \colhead{Program} & \multicolumn{2}{c}{Channel Width} & \colhead{Median} & \colhead{$t_{\rm int}$} & \colhead{Median Beam} & \colhead{PA} & \colhead{MJD\tablenotemark{a}} & \colhead{rms\tablenotemark{b}} \\[-4pt] \cline{4-5} 
  & & & \colhead{Obs.} & \colhead{Sm.} & \colhead{Velocity}\\[-3pt]
  & (GHz) & & (MHz) & (MHz) & (km~s$^{-1}$) & (s) & (arcsec) & ($^\circ$) & &  (mJy) }
  \startdata
  L & 1.008--2.032 & 14A-231 & 1  & 10  & 2000 & 10591 & \ $2.2\times1.1$\tablenotemark{c} & 1 & 56749 & 0.33\tablenotemark{d}\\
  S & 2.157--3.961 & BP198 & 0.5 & 12  & 1172 & 26361 & \ $3.1\times1.2$\tablenotemark{c} & 3 & 57577--57580 & 0.14\tablenotemark{d,e}\\
  C & 4.487--6.511  & 14A-231 & 2 & 14 & 763 & 10591  & $0.61\times0.28$ & $-$3 & 56749 & 0.099\\
  X & 7.987--10.011  & 14A-231 & 2 & 18 & 600 & 19148 & $0.34\times0.17$ & $-$1 & 56718 &  0.026\tablenotemark{f}\\
  X & 8.007--11.991  & 15A-418 & 2 & 18 & 539 & 27290 & $0.68\times0.38$ & $-$53 & 57167--57173 &  0.098\\
Ku & 12.038--13.060 & 12A-339 & 2 & 20 & 463 & 4086 & $0.84\times0.47$ & 3 & 56141--56143 &   0.066\\
Ku & 12.988--15.012 & 14A-231 & 2 & 20 & 428 & 16156 & $0.24\times0.11$ & $-$1 & 56726 &   0.027\\ 
Ku & 16.951--17.961 & 12A-339 & 2 & 22 & 372 & 4086 & $0.64\times0.36$ & $-$3 & 56141--56143 &   0.082\\
K  & 18.875--19.511 & 12A-339 & 2 & 22 & 349 & 8499 & $0.58\times0.28$ & 1 & 56065--56143 &   0.092\\
K  & 25.501--26.511 & 12A-339 & 2 & 24 & 285 & 8499 & $0.43\times0.20$ & 1 & 56065--56143 &  0.085\\ 
Ka & 26.975--27.863 & 12A-339 & 2 & 26 & 275 & 8499 & $0.37\times0.17$ & 1 & 56065--56143 &  0.17\\
Ka & 30.476--32.524 & 14A-232 & 2 & 26 & 247 & 17053  & $0.101\times0.050$ & $-$3 & 56725 &  0.100\\
Ka & 32.476--34.524 & 14A-232 & 2 & 26 & 233 & 17053  & $0.094\times0.046$ & $-$3 & 56725 &  0.116\\
Ka & 34.476--36.524 & 14A-232 & 2 & 28 & 236 & 17053  & $0.089\times0.045$ & $-$4 & 56725 &  0.165\\
Ka & 36.476--38.524 & 14A-232 & 2 & 28 & 224 & 17053  & $0.084\times0.042$ & $-$3 & 56725 &  0.152\\
Ka & 37.493--38.500 & 12A-339 & 2 & 28 & 221 & 8499 & $0.28\times0.13$ & 1 & 56065--56143 &  0.10\\
Q  & 39.300--40.052 & 12A-339 & 2 & 28 & 215 & 4413 & $0.37\times0.27$ & 36 & 56065--56123 &  0.22\\
\enddata
\tablenotetext{a}{Modified Julian Date. Ranges indicate the span of dates included in a program. }
\tablenotetext{b}{The spectral rms noise in Gaussian-smoothed channels of width $\Delta f$ (Column 5 and Equation \ref{eqn:df}).}
\tablenotetext{c}{The quoted beam is the continuum beam; the synthesized beam in the spectral cube is highly variable due to RFI and the large fractional bandwidth.} 
\tablenotetext{d}{The rms noise includes residual unmitigated RFI.}
\tablenotetext{e}{The rms noise is measured in the 3--4~GHz spectrum.}
\tablenotetext{f}{The rms noise omits the central RFI feature and band edges.}
\tablecomments{Results for programs 14A-231 and 14A-232 were published in \citet{darling2020} and are reproduced here for completeness.}
\end{deluxetable*}

\section{\label{sec:obs}Observations}

\citet{darling2020} used archival VLA observations of Sgr A* and/or PSR J1745$-$2900 (both are present in every primary beam),
that have the highest angular resolution (A-array) in order to separate the magnetar from Sgr A* and to resolve out extended spectral line-emitting
Galactic Center gas and extended continuum emission.
The work presented here adds observations that are sub-optimal (B- and C-array) but which still enable the angular separation of
Sgr A* from PSR J1745$-$2900.  Unlike for the observations presented in \citet{darling2020}, the PSR J1745$-$2900 radio continuum cannot
be separated from the surrounding emission in these lower angular resolution data.
Based on our previous confirmation of the astrometry of PSR J1745$-$2900 conducted by \citet{bower2015}, we are confident in our ability to extract a spectrum of  PSR J1745$-$2900 from interferometric image cubes.  

VLA observations of Sgr A* and PSR J1745$-$2900 were selected
(1) to maximize on-target integration time, (2) with adequate angular resolution to separate PSR J1745$-$2900
from Sgr A* (1.7'' in both coordinates), (3) to maximize total bandwidth, and (4) to adequately sample the expected emission line bandwidth.
In addition to the VLA A-configuration programs 14A-231 and 14A-232 presented in \citet{darling2020},
programs 12A-339, BP198, and 15A-418 meet these criteria and are analyzed in this Letter (Table \ref{tab:obs}).

Observing sessions used  J1331+3030 (3C286) and J0137+331 (3C48) for flux calibration, 3C286, 3C48,
and J1733$-$1304 for bandpass calibration, and J1744$-$3116, J1745$-$283, and J1751$-$2524 for complex gain calibration.
Right- and left-circular polarizations were combined to form Stokes-I spectral cubes.  
Bandwidths of spectral windows were either 32 MHz or 128 MHz, subdivided into 0.5 or 2 MHz channels, respectively, and grouped into
2--4 overlapping basebands.  All programs used 8-bit sampling except for 15A-418, which used 3-bit sampling.  Correlator dump times were 1--5 s.

\section{\label{sec:reduc}Data Reduction}

We used CASA\footnote{McMullin, J. P., Waters, B., Schiebel, D., Young, W., \& Golap, K. 2007, Astronomical Data Analysis Software and Systems XVI (ASP Conf. Ser. 376), ed. R. A. Shaw, F. Hill, \& D. J. Bell (San Francisco, CA: ASP), 127} for interferometric data reduction.
Data were flagged and calibrated (flux, delay, atmospheric transmission, complex bandpass, and complex gain).  After applying calibration
to the target field, we did in-beam phase self-calibration on the Sgr A* continuum (0.8--1.7 Jy from S- to Q-band).

We imaged the continuum in the target field and fit a 2D Gaussian to Sgr A* to set the origin for relative astrometry.  We locate PSR J1745$-$2900 using the bootstrap proper motion solution obtained by \citet{bower2015}.  Offsets were consistent
between epochs in each band and between bands and were consistent with the observed continuum position of the magnetar
detected in programs 14A-231 and 14A-232 \citep{darling2020}.  

After linear continuum subtraction in {\it uv} space, we formed spectral image cubes and cleaned these down to
five times the theoretical noise.
Sgr A* shows narrow-band spectral structure after the continuum subtraction, particularly in X-band where we see a comb of radio recombination lines (RRLs) in emission, but also due to spectral window edge effects.
The RRLs are presumably mildly stimulated.  We also see extended RRL emission in many spectral
cubes due to the low angular resolution, and we see RRL emission toward the magnetar in some bands (see below).
Sgr A* sidelobes are cleaned during cube deconvolution and do not significantly contaminate the magnetar spectrum, and
the magnetar spectra typically reach the theoretical noise.   Synthesized beams
vary across each spectral cube due to the natural frequency-dependent angular resolution of the interferometer and due to
data flagging and RFI.  

We extract the magnetar spectrum over the 2D Gaussian beam and correct for the beam size for each channel
in order to capture the total point source flux density.  The spectral noise varies channel-to-channel, which can create
false peaks in the magnetar spectrum.  To assess significance of spectral features, we form a noise spectrum using a measurement
of the sky rms noise in each channel.   The overall noise spectrum is scaled to the spectral noise of PSR J1745$-$2900, which typically
differ by a few percent.

For single-channel detection, we need to smooth spectra to the expected axion-photon conversion line width, but
there is theoretical disagreement about the expected bandwidth of the emission line.  \citet{hook2018} make
a conservation of energy argument to derive a fractional bandwidth that depends quadratically on the axion velocity dispersion $v_0$:
$\Delta f/f = (v_0/c)^2$ \citep[contrary to the intuitive expectation for the line width to reflect the velocity
dispersion as a Doppler shift;][]{huang2018}.  \citet{battye2020}, however, suggest that the line width is dominated by the
neutron star's spinning magnetosphere.  We adopt this spinning mirror model which produces, on average, a bandwidth
$\Delta f/f \simeq \Omega\, r_c /c$, where $\Omega$ is the rotation angular frequency
   %    $\varepsilon$ is the eccentricity of the elliptical critical surface,
and $r_c$ is the axion-photon conversion radius.
\citet{hook2018} show that $r_c$ depends on the neutron star's radius $r_0$, magnetic field $B_0$, angular frequency,
polar orientation angle $\theta$, and magnetic axis offset angle $\theta_m$:
\begin{eqnarray}\label{eqn:rc}
  r_c = 224~{\rm km} \times \left| 3 \cos\theta\ \hat{m}\cdot\hat{r} - \cos\theta_m\right|^{1/3} \times \nonumber \\
    ~\left(r_0 \over 10~{\rm km}\right) \left[ {B_0 \over 10^{14}~{\rm G}} {1\over 2\pi} {\Omega \over 1~{\rm Hz}} \left(4.1~\mu{\rm eV} \over m_a c^2\right)^2\right]^{1/3}
\end{eqnarray}
where $\hat{m}\cdot\hat{r} = \cos\theta_m \cos\theta + \sin\theta_m \sin\theta \cos\Omega t$.
For now, we assume that $\theta = \pi/2$ and $\theta_m = 0$ (we deal appropriately with these angles in Section \ref{sec:analysis}) to
obtain the expected line width:
\begin{equation}\label{eqn:df}
  \Delta f = 3.6~{\rm MHz} \left(\Omega \over 1~{\rm Hz}\right)^{4/3} \left(m_a c^2 \over 4.1~\mu{\rm eV}\right)^{1/3} \left(B_0 \over 10^{14}~{\rm G}\right)^{1/3}
\end{equation}
(4.1~$\mu$eV corresponds to 1 GHz as observed).  PSR J1745$-$2900 has a  3.76~s rotation period \citep{kennea2013} and a magnetic field
of $1.6\times10^{14}$~G \citep{mori2013}.  The expected axion-photon conversion line width is
thus $\Delta f = 8.3~{\rm MHz} \times (m_a c^2 / 4.1~\mu{\rm eV})^{1/3}$.
This corresponds to 2500~km~s$^{-1}$ at 1~GHz and 215~km~s$^{-1}$ at 40~GHz,
which is generally broader than the expected dark matter dispersion, $\sim$300~km~s$^{-1}$, except at the highest observed frequencies.

Spectra have flagged channels due to RFI and due to spectral window edges that lack the signal-to-noise for
calibration.  Flagged channels are much narrower than the expected line width, so we interpolate across these channels
when smoothing using a Gaussian kernel \citep{astropy:2018}.  In some bands, such as L, S, and K, entire spectral windows can be flagged
and all information is lost.  

RRL emission lines are seen toward the magnetar in X-band (15A-418), Ku 12--13 GHz, Ku 17--18~GHz, and K 26~GHz.
% only one session in latter two bands
To remove RRLs, two 2~MHz channels are flagged per line, and subsequent smoothing interpolates
across the flagged channels (the expected axion conversion line width is 18 MHz in X-band and 24 MHz at 26~GHz).

We combine spectra obtained from multiple observing sessions using an error-weighted average, and the sky (noise) spectra are
combined in quadrature.  When different observing programs overlap in frequency, we select the lowest-noise observation.  This is
effectively the same as combining overlapping spectra in quadrature because the less sensitive spectra contribute negligibly to
an error-weighted mean.  

Table \ref{tab:obs} lists synthesized beam parameters,  channel widths, and spectral rms noise values.  Appendix \ref{sec:appendix}
presents the new magnetar flux, noise, and signal-to-noise spectra.

\begin{figure}[t]
\begin{centering}
  \includegraphics[scale=0.25,trim=20 20 20 0,clip=true]{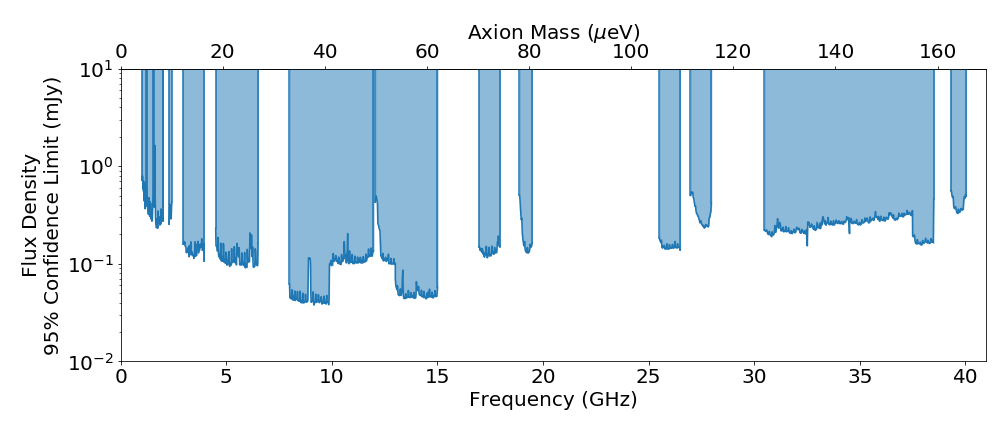}
  \includegraphics[scale=0.25,trim=20 20 20 0,clip=true]{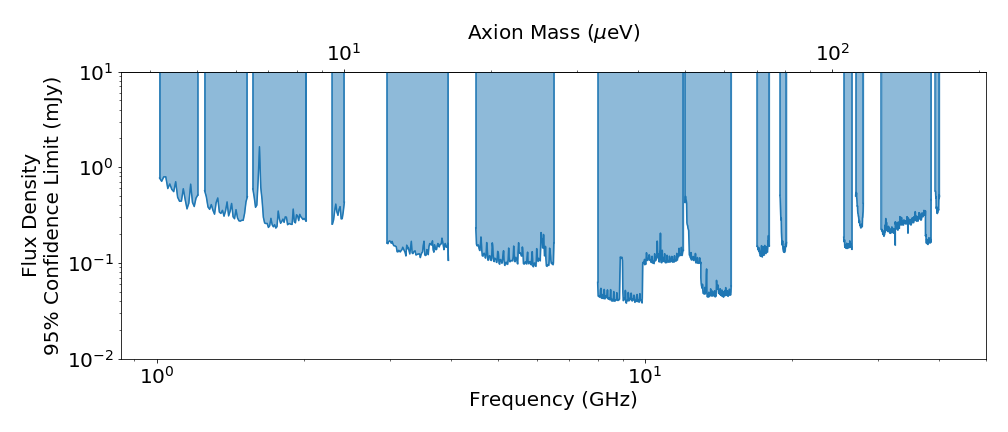}
\caption{%\footnotesize
  95\% confidence limits on axion-photon conversion line flux density.  
  We present both linear (top) and logarithmic (bottom) scales for ease of comparison to other work.}
\label{fig:flux}
\end{centering}
\end{figure}

\section{\label{sec:results}Results}

After smoothing to the expected axion-photon conversion line width, the new spectra show no significant ($>3\sigma$) single-channel
emission features (Appendix \ref{sec:appendix}).  The exception is a 3.2$\sigma$ channel at 2.34~GHz (9.67~$\mu$eV) in an RFI-affected
region of S-band \citep[there also two channels at 3.1$\sigma$ and 3.5$\sigma$ in previous Ka-band spectra;][]{darling2020}.

We obtain single-channel 95\% confidence limit flux density spectra from the sky noise spectra.  Figure \ref{fig:flux}
shows the combined limits from this work and \citet{darling2020}.  These limits do depend on the assumed line width (Equation
\ref{eqn:df}), which depends on the magnetar model, but these limits may be scaled as needed for magnetar models not treated in this Letter.

\section{\label{sec:analysis}Analysis}

Translating spectral flux density limits into limits on the axion-photon coupling $g_{a\gamma\gamma}$ depends on the
axion-photon conversion in the magnetar magnetosphere and on the density of dark matter in the Galactic Center.  Both
of these rely on as-yet incompletely constrained models.

\subsection{\label{subsec:magnetar}The Magnetar Model}

We adopt the \citet{hook2018} axion-photon conversion model for the magnetar magnetosphere, which is based on a
variant of the \citet{goldreich1969} model (but note that there is substantial disagreement about the signal bandwidth
and radiated power in the literature; e.g., \citet{hook2018,huang2018,leroy2020,battye2020}).  We modify this model
for the bandwidth adopted above (Equation \ref{eqn:df}) to obtain an expression for the observed
flux density in the axion-photon conversion emission line that depends on the magnetar properties, distance, viewing angle
$\theta$, and local dark matter density \citep[see][Equation 3]{darling2020}.

\citet{darling2020} shows that given a flux density limit spectrum, one can produce a limit on
$g_{a\gamma\gamma}$ as a function of $m_a$ that depends on the dark matter velocity dispersion $v_0$, the
dark matter density $\rho_\infty$, and a time-dependent angular term involving $\theta$, $\theta_m$, and the axion velocity
at the conversion point, $v_c^2 \simeq 2G M_{NS}/r_c$ \citep{battye2020}.
For PSR J1745$-$2900 specifically, assuming a magnetar radius of 10~km, mass of 1~M$_\odot$, and distance of 8.2~kpc,
we obtain
\begin{widetext}
\begin{eqnarray}\label{eqn:g_agg} 
  g_{a\gamma\gamma} =  3 \times 10^{-11}\ {\rm GeV}^{-1}\
  \left(S_\nu \over 10\ \mu{\rm Jy}\right)^{1/2}\ 
  \left(m_a \over 1\ {\rm GHz}\right)^{-2/3} %\nonumber \\
  \left(v_0\over 200\ {\rm km~s}^{-1}\right)^{1/2}\ \nonumber \\
  \times
  \left(\rho_\infty \over 6.5\times10^4\  {\rm GeV~cm}^{-3}\right)^{-1/2} \ %\nonumber \\
  \left({3 (\hat{m}\cdot\hat{r})^2 + 1 \over |3\cos\theta\ \hat{m}\cdot\hat{r}-\cos\theta_m|^{4/3}} {v_c\over c}\right)^{-1/2}.
\end{eqnarray}
\end{widetext}
The angular term relies on the unknown viewing and magnetic field misalignment angles and is time-dependent.
Axion-photon conversion also relies on the conversion radius being outside the magnetar surface  ($r_c > r_0$; see Equation \ref{eqn:rc}),
which is axion mass-dependent.   For a given $(\theta,\theta_m)$ pair, there may be parts of the magnetar rotation
period that do not radiate.
Since the modulation time is much less than the integration time of the observations, we average the expected signal
over the period of the magnetar separately for each frequency channel for each $(\theta,\theta_m)$ pair, to form time-integrated
flux density spectra.  We then marginalize over all $(\theta,\theta_m)$ to obtain a limit spectrum on $g_{a\gamma\gamma}$ given
a dark matter density and velocity dispersion (see below).

The ray-tracing performed by \citet{leroy2020} suggests that this analytic treatment is conservative and that axion-photon
conversion can occur over a larger parameter space.
Nonetheless, the signal losses caused by angles where $r_c$ is always less than $r_0$ in this analytic treatment
are a small fraction of the parameter space: at 10~GHz, 0.08\% of all possible $(\theta,\theta_m)$ always have $r_c < r_0$.
This grows with frequency, and at 40~GHz the fraction of orientations with no emission rises to 6.6\%.

\begin{figure*}
\begin{centering}
  \includegraphics[scale=0.52,trim=20 20 20 0,clip=true]{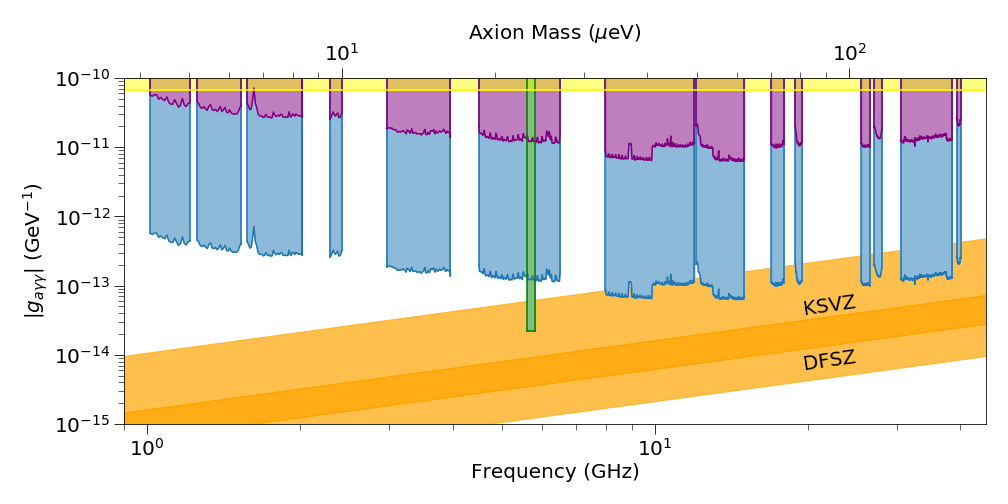}
\caption{%\footnotesize
  95\% confidence limits on $|g_{a\gamma\gamma}|$ for the NFW model prediction of the Galactic Center dark matter energy density (purple, upper)
  and the same NFW model plus a maximal central 100~pc dark matter spike (blue, lower).
  The HAYSTAC limit \citep[green][]{Zhong2018} has been scaled from a local axion density of 0.45~GeV~cm$^{-3}$  to 0.38~GeV~cm$^{-3}$.
  The yellow bar shows the CAST 95\% confidence limit obtained from a search for solar axions \citep{Anastassopoulos2017}.
  The orange bands indicate the range of possible QCD axion models \citep{Luzio2017}, which include the canonical KSVZ and DFSZ models
  \citep{Kim1979,Shifman1980,Dine1981,Zhitnitsky1980}.  % [UF/RBF loci?]
}
\label{fig:limits}
\end{centering}
\end{figure*}

\begin{deluxetable}{ccc}
  \tablecaption{\label{tab:g_agg}  Limits on the Axion-Photon Coupling $g_{a\gamma\gamma}$}
%\begin{ruledtabular}
%\begin{tabular}{ccc}
  \tablehead{ \colhead{Axion Mass}  & \multicolumn{2}{c}{Median $|g_{a\gamma\gamma}|$ 95\% Confidence Limits} \\[2pt] \cline{2-3} \\[-10pt] 
     & NFW Profile & DM Spike \\
       ($\mu$eV)  & (GeV$^{-1}$) & (GeV$^{-1}$) }
    \startdata
   4.2--8.4\tablenotemark{a} & $3.4 \times 10^{-11}$ &  $3.4 \times 10^{-13}$ \\
   8.9--10.0 &  $2.9 \times 10^{-11}$ &   $2.9 \times 10^{-13}$\\
  12.3--16.4 &  $1.7 \times 10^{-11}$ &   $1.7 \times 10^{-13}$\\
  18.6--26.9 &  $1.3 \times 10^{-11}$ &   $1.3 \times 10^{-13}$\\
  33.0--41.3 &  $6.9 \times 10^{-12}$ &   $7.0 \times 10^{-14}$\\
  41.3--49.6 &  $1.1 \times 10^{-11}$ &   $1.1 \times 10^{-13}$\\
  49.8--53.8 &  $1.0 \times 10^{-11}$ &   $1.0 \times 10^{-13}$\\
  53.8--62.1 &  $6.5 \times 10^{-12}$ &   $6.5 \times 10^{-14}$\\
  70.1--74.3 &  $1.0 \times 10^{-11}$ &   $1.0 \times 10^{-13}$\\
  78.1--80.7 &  $1.1 \times 10^{-11}$ &   $1.2 \times 10^{-13}$\\
  105.5--109.6 &  $1.0 \times 10^{-11}$ &   $1.0 \times 10^{-13}$\\
  111.6--115.2 &  $1.5 \times 10^{-11}$ &   $1.5 \times 10^{-13}$\\
  126.0--155.1 & $1.4 \times 10^{-11}$ &   $1.4 \times 10^{-13}$\\
  155.1--159.3 & $1.3 \times 10^{-11}$ &   $1.3 \times 10^{-13}$\\
  162.5--165.6 & $2.2 \times 10^{-11}$ &   $2.2 \times 10^{-13}$\\
\enddata
   %  \end{tabular}
%\end{ruledtabular}
\tablenotetext{a}{There are gaps in the coverage of this mass range (see Figure \ref{fig:limits}).}
\end{deluxetable}

\subsection{\label{subsec:dm} Dark Matter Models}

The remaining unknowns in Equation \ref{eqn:g_agg} are the dark matter density $\rho_\infty$ and velocity dispersion $v_0$
at the location of the magnetar.  The dark matter contribution in the Galactic Center has
not been measured, so one must employ model-based interpolation.  Following \citet{hook2018}, we adopt two models that
roughly bracket the possible dark matter density (unless the dark matter distribution is cored; a multi-kpc central
core is disfavored by observations \citep[e.g.][]{hooper2017}, and there are reasons to believe that baryon contraction
has occurred \citep{cautun2020}, but the existence of a cusp or core in the inner kpc remains observationally
untested):  a Navarro-Frenk-White
\citep[NFW;][]{NFW} dark matter profile and the same model plus a maximal dark matter cusp.   For both models, we assume
$v_0 = 300$~km~s$^{-1}$ and a 0.1~pc separation between PSR J1745$-$2900 and Sgr A* (we identify Sgr A* with the center
of the dark matter distribution).  The 0.1~pc separation between PSR J1745$-$2900 and Sgr A* is projected, but absent an
acceleration measurement one cannot know their true physical separation \citep{bower2015}. 

For the NFW dark matter profile, we adopt the \citet{McMillan2017} fit
with scale index $\gamma=1$, scale radius $r_s = 18.6$~kpc, local dark matter energy
density $\rho_\odot = 0.38$~GeV~cm$^{-3}$, and Galactic Center distance $R_0 = 8.2$~kpc, which agrees with the measurement
by \citet{Abuter2019} of the orbit of the star S2 about Sgr A*.  This model predicts a dark matter energy density of
$6.5\times10^4$~GeV~cm$^{-3}$ at 0.1~pc.

The dark matter cusp model adds a spike to the NFW model with scale $R_{sp}  = 100$~pc and scale index $\gamma_{sp} = 7/3$.
This is the maximal dark matter spike corresponding to a 99.7\% upper limit derived from a lack of deviations of the S2 star from
a black hole-only orbit about Sgr A* \citep{Lacroix2018}.  The maximal dark matter energy density encountered by the magnetar
is thus $6.4\times10^8$~GeV~cm$^{-3}$, a factor of $10^4$ larger than the NFW-only model.  This enhanced dark matter
density corresponds to a 100-fold smaller constraint on $g_{a\gamma\gamma}$.

We present band-median 95\% confidence limits on $|g_{a\gamma\gamma}|$ for each dark matter model in Table \ref{tab:g_agg}.
Figure \ref{fig:limits} shows the limit spectra spanning 62\% of the 1--40~GHz (4.2--165.6~$\mu$eV) range,
previous limits from CAST and HAYSTAC \citep{Anastassopoulos2017,Zhong2018}, and the family of theoretical axion models
\citep{Luzio2017}.  Limits obtained from the NFW model exclude $|g_{a\gamma\gamma}| \gtrsim $ 6--34 $\times10^{-12}$~GeV$^{-1}$,
which is 1.5--3.5 dex above the strongest-coupling theoretical prediction.  
The maximal dark matter spike model limits do, however, exclude portions of theoretical parameter space for $m_a = 33.0$--165.6~$\mu$eV.
The canonical KSVZ or DFSZ models are not excluded \citep{Kim1979,Shifman1980,Dine1981,Zhitnitsky1980}.

\section{\label{sec:discussion}Discussion}

The limits on $g_{a\gamma\gamma}$ presented here for the NFW profile are conservative compared to the \citet{hook2018} predictions.
This is due to the choice of a spinning-mirror bandwidth that seems more physically plausible \citep{battye2020}. This bandwidth is
$\mathcal{O}(v_0/c)$, roughly $10^3$ times larger than the \citet{hook2018} $\mathcal{O}(v_0/c)^2$ bandwidth.  This is a factor of $\sim$300
in $g_{a\gamma\gamma}$.  A better treatment of this issue will require axion-photon conversion ray tracing as proposed by \citet{leroy2020}.  

Our limits may also be conservative because resonant axion-photon conversion may be stimulated by the local photon occupation number, which
would boost any signal thereby improving constraints on  $g_{a\gamma\gamma}$ \citep{Caputo2019}.  It seems likely that stimulated emission
would be particularly important in the Galactic Center photon bath, but it may also arise naturally from the magnetar itself.  This effect and
numerical ray-tracing may significantly improve the constraints on $g_{a\gamma\gamma}$ based on the current observations alone.  

As Figure \ref{fig:limits} shows, the highly uncertain dark matter energy density in the inner parsec allows a large range of possible constraints
on axion parameter space.  Moreover, if the central dark matter is cored with a fixed NFW density of 12~GeV~cm$^{-3}$ inward of 500 pc ,
then the limits on $g_{a\gamma\gamma}$ are degraded by 2 dex and lie above the CAST limits.
We look forward to observational measurements of or constraints on the Galactic Center dark matter
encountered by PSR~J1745$-$2900 based on stellar and gas dynamics.

\section{\label{sec:conclusions}Conclusions}

We have expanded the axion mass range searched for the axion-photon conversion signal originating from the magnetosphere of the
Galactic Center magnetar PSR~J1745$-$2900.  New limits span 62\% of the 4.2--165.6~$\mu$eV (1--40~GHz) axion mass range, excluding
at 95\% confidence $g_{a\gamma\gamma} > $ 6--34 $\times10^{-12}$~GeV$^{-1}$ if the dark matter energy density follows a generic
NFW profile at the Galactic Center.  For a maximal dark matter spike, the limit reduces to $g_{a\gamma\gamma} > $ 6--34 $\times10^{-14}$~GeV$^{-1}$,
which excludes some possible axion models for $m_a > 33$~$\mu$eV.  

This work gets close to exhausting the appropriate data in the VLA archive.  Lower resolution interferometric observations
cannot separate the magnetar from the Sgr A* continuum, and we demonstrate how the extended Galactic Center continuum and line emission
impairs the identification of the magnetar continuum and impacts the low angular resolution spectrum (particularly the RRL emission).
Future observations designed to fill in the axion mass coverage or to increase sensitivity should use sub-arcsec resolution arrays.  But high-resolution
observations will exclude any axion-photon conversion signal that may arise from an extended population of Galactic Center neutron stars
\citep{Safdi2019}.  

It is unclear whether an axion-photon conversion signal will be pulsed.  Analytic models suggest that it should be \citep[e.g.][]{hook2018},
but detailed ray-tracing and magnetosphere simulation are needed \citep{leroy2020}.  If emission is pulsed, future observations could
in principle increase signal-to-noise by observing spectra in a gated pulsar mode.

\acknowledgments
  We thank the operations, observing, archive, and computing staff at the NRAO who made this work possible.  We also thank
  Konrad Lehnert, Marco Chianese, Andrea Caputo, and Richard Battye for helpful discussions.
  This research made use of CASA \citep{CASA}, NumPy \citep{NumPy}, Matplotlib \citep{Matplotlib}, and
  Astropy\footnote{\url{http://www.astropy.org}}, a community-developed core Python package for Astronomy \citep{astropy:2013, astropy:2018}.

  \facility{VLA}
  \software{CASA (McMullin et al. 2007), astropy (The Astropy Collaboration 2013, 2018), NumPy \citep{NumPy}, Matplotlib \citep{Matplotlib} }

\bibliography{ms}% Produces the bibliography via BibTeX.

\providecommand{\noopsort}[1]{}\providecommand{\singleletter}[1]{#1}%
\begin{thebibliography}{}
\expandafter\ifx\csname natexlab\endcsname\relax\def\natexlab#1{#1}\fi

\bibitem[{{Abbott} \& {Sikivie}(1983)}]{abbott1983}
{Abbott}, L.~F., \& {Sikivie}, P. 1983, Physics Letters B, 120, 133

\bibitem[{Abuter {et~al.}(2019)Abuter, Amorim, Baub{\"{o}}ck, Berger, Bonnet,
  Brandner, Cl{\'{e}}net, {Coud{\'{e}} du Foresto}, de~Zeeuw, Dexter, Duvert,
  Eckart, Eisenhauer, {F{\"{o}}rster Schreiber}, Garcia, Gao, Gendron, Genzel,
  Gerhard, Gillessen, Habibi, Haubois, Henning, Hippler, Horrobin,
  Jim{\'{e}}nez-Rosales, Jocou, Kervella, Lacour, Lapeyr{\`{e}}re, {Le
  Bouquin}, L{\'{e}}na, Ott, Paumard, Perraut, Perrin, Pfuhl, Rabien,
  {Rodriguez Coira}, Rousset, Scheithauer, Sternberg, Straub, Straubmeier,
  Sturm, Tacconi, Vincent, von Fellenberg, Waisberg, Widmann, Wieprecht,
  Wiezorrek, Woillez, \& Yazici}]{Abuter2019}
Abuter, R., Amorim, A., Baub{\"{o}}ck, M., {et~al.} 2019, Astronomy {\&}
  Astrophysics, 625, L10

\bibitem[{Anastassopoulos {et~al.}(2017)Anastassopoulos, Aune, Barth, Belov,
  Br{\"{a}}uninger, Cantatore, Carmona, Castel, Cetin, Christensen, Collar,
  Dafni, Davenport, Decker, Dermenev, Desch, Eleftheriadis, Fanourakis,
  Ferrer-Ribas, Fischer, Garc{\'{i}}a, Gardikiotis, Garza, Gazis, Geralis,
  Giomataris, Gninenko, Hailey, Hasinoff, Hoffmann, Iguaz, Irastorza, Jakobsen,
  Jacoby, Jakovcic, Kaminski, Karuza, Kralj, Krcmar, Kostoglou, Krieger, Lakic,
  Laurent, Liolios, Ljubicic, Luz{\'{o}}n, Maroudas, Miceli, Neff, Ortega,
  Papaevangelou, Paraschou, Pivovaroff, Raffelt, Rosu, Ruz, Ch{\'{o}}liz,
  Savvidis, Schmidt, Semertzidis, Solanki, Stewart, Vafeiadis, Vogel, Yildiz,
  \& Zioutas}]{Anastassopoulos2017}
Anastassopoulos, V., Aune, S., Barth, K., {et~al.} 2017, Nature Physics, 13,
  584

\bibitem[{Arik \& et~al.(2014)}]{arik2014}
Arik, M., \& et~al. 2014, Phys.\ Rev.\ Lett., 112, 091302

\bibitem[{Arik \& et~al.(2015)}]{arik2015}
---. 2015, Phys.\ Rev.\ D, 92, 021101

\bibitem[{{Astropy Collaboration} {et~al.}(2013){Astropy Collaboration},
  {Robitaille}, {Tollerud}, {Greenfield}, {Droettboom}, {Bray}, {Aldcroft},
  {Davis}, {Ginsburg}, {Price-Whelan}, {Kerzendorf}, {Conley}, {Crighton},
  {Barbary}, {Muna}, {Ferguson}, {Grollier}, {Parikh}, {Nair}, {Unther},
  {Deil}, {Woillez}, {Conseil}, {Kramer}, {Turner}, {Singer}, {Fox}, {Weaver},
  {Zabalza}, {Edwards}, {Azalee Bostroem}, {Burke}, {Casey}, {Crawford},
  {Dencheva}, {Ely}, {Jenness}, {Labrie}, {Lim}, {Pierfederici}, {Pontzen},
  {Ptak}, {Refsdal}, {Servillat}, \& {Streicher}}]{astropy:2013}
{Astropy Collaboration}, {Robitaille}, T.~P., {Tollerud}, E.~J., {et~al.} 2013,
  A\&A, 558, A33

\bibitem[{Asztalos \& et~al.(2001)}]{asztalos2001}
Asztalos, S.~J., \& et~al. 2001, Phys.\ Rev.\ D, 64, 092003

\bibitem[{Asztalos \& et~al.(2010)}]{asztalos2010}
---. 2010, Phys.\ Rev.\ Lett., 104, 041301

\bibitem[{{Battye} {et~al.}(2020){Battye}, {Garbrecht}, {McDonald}, {Pace}, \&
  {Srinivasan}}]{battye2020}
{Battye}, R.~A., {Garbrecht}, B., {McDonald}, J.~I., {Pace}, F., \&
  {Srinivasan}, S. 2020, \prd, 102, 023504

\bibitem[{Bower {et~al.}(2015)Bower, Deller, Demorest, Brunthaler, Falcke,
  Moscibrodzka, O'Leary, Eatough, Kramer, Lee, Spitler, Desvignes, Rushton,
  Doeleman, \& Reid}]{bower2015}
Bower, G.~C., Deller, A., Demorest, P., {et~al.} 2015, The Astrophysical
  Journal, 798, 120

\bibitem[{Brubaker \& et~al.(2017)}]{brubaker2017}
Brubaker, B.~M., \& et~al. 2017, Phys.\ Rev.\ Lett., 118, 061302

\bibitem[{Caputo {et~al.}(2019)Caputo, Regis, Taoso, \& Witte}]{Caputo2019}
Caputo, A., Regis, M., Taoso, M., \& Witte, S.~J. 2019, Journal of Cosmology
  and Astroparticle Physics, 2019, arXiv:1811.08436

\bibitem[{{Cautun} {et~al.}(2020){Cautun}, {Ben{\'\i}tez-Llambay}, {Deason},
  {Frenk}, {Fattahi}, {G{\'o}mez}, {Grand}, {Oman}, {Navarro}, \&
  {Simpson}}]{cautun2020}
{Cautun}, M., {Ben{\'\i}tez-Llambay}, A., {Deason}, A.~J., {et~al.} 2020,
  MNRAS, 494, 4291

\bibitem[{{Co} \& {Harigaya}(2019)}]{Co2020}
{Co}, R.~T., \& {Harigaya}, K. 2019, arXiv e-prints, arXiv:1910.02080

\bibitem[{{Darling}(2020)}]{darling2020}
{Darling}, J. 2020, \prl, in press, arXiv:2008.01877

\bibitem[{{Day} \& {McDonald}(2019)}]{day2019}
{Day}, F.~V., \& {McDonald}, J.~I. 2019, JCAP, 2019, 051

\bibitem[{{Di Luzio} {et~al.}(2017){Di Luzio}, Mescia, \& Nardi}]{Luzio2017}
{Di Luzio}, L., Mescia, F., \& Nardi, E. 2017, Physical Review Letters, 118,
  arXiv:1610.07593

\bibitem[{Dine \& Fischler(1983)}]{dine1983}
Dine, M., \& Fischler, W. 1983, Physics Letters B, 120, 137

\bibitem[{{Dine} {et~al.}(1981){Dine}, {Fischler}, \& {Srednicki}}]{Dine1981}
{Dine}, M., {Fischler}, W., \& {Srednicki}, M. 1981, Physics Letters B, 104,
  199

\bibitem[{{Edwards} {et~al.}(2020){Edwards}, {Chianese}, {Kavanagh},
  {Nissanke}, \& {Weniger}}]{edwards2020}
{Edwards}, T. D.~P., {Chianese}, M., {Kavanagh}, B.~J., {Nissanke}, S.~M., \&
  {Weniger}, C. 2020, \prl, 124, 161101

\bibitem[{Goldreich \& Julian(1969)}]{goldreich1969}
Goldreich, P., \& Julian, W.~H. 1969, ApJ, 157, 869

\bibitem[{Hook {et~al.}(2018)Hook, Kahn, Safdi, \& Sun}]{hook2018}
Hook, A., Kahn, Y., Safdi, B., \& Sun, Z. 2018, Phys.\ Rev.\ Lett., 121, 241102

\bibitem[{Hooper(2017)}]{hooper2017}
Hooper, D. 2017, Physics of the Dark Universe, 15, 53

\bibitem[{Huang {et~al.}(2018)Huang, Kadota, Sekiguchi, \& Tashiro}]{huang2018}
Huang, F.~P., Kadota, K., Sekiguchi, T., \& Tashiro, H. 2018, Physical Review
  D, 97, 123001

\bibitem[{{Hunter}(2007)}]{Matplotlib}
{Hunter}, J.~D. 2007, Computing in Science Engineering, 9, 90

\bibitem[{Kennea \& et~al.(2013)}]{kennea2013}
Kennea, J.~A., \& et~al. 2013, ApJ, 770, L24

\bibitem[{{Kim}(1979)}]{Kim1979}
{Kim}, J.~E. 1979, \prl, 43, 103

\bibitem[{Lacroix(2018)}]{Lacroix2018}
Lacroix, T. 2018, Astronomy and Astrophysics, 619, 46

\bibitem[{{Leroy} {et~al.}(2020){Leroy}, {Chianese}, {Edwards}, \&
  {Weniger}}]{leroy2020}
{Leroy}, M., {Chianese}, M., {Edwards}, T. D.~P., \& {Weniger}, C. 2020, \prd,
  101, 123003

\bibitem[{{McMillan}(2017)}]{McMillan2017}
{McMillan}, P.~J. 2017, MNRAS, 465, 76

\bibitem[{{McMullin} {et~al.}(2007){McMullin}, {Waters}, {Schiebel}, {Young},
  \& {Golap}}]{CASA}
{McMullin}, J.~P., {Waters}, B., {Schiebel}, D., {Young}, W., \& {Golap}, K.
  2007, 127

\bibitem[{Mori \& et~al.(2013)}]{mori2013}
Mori, K., \& et~al. 2013, ApJ, 770, L23

\bibitem[{{Mukherjee} {et~al.}(2020){Mukherjee}, {Spergel}, {Khatri}, \& {Wand
  elt}}]{mukherjee2020}
{Mukherjee}, S., {Spergel}, D.~N., {Khatri}, R., \& {Wand elt}, B.~D. 2020,
  JCAP, 2020, 032

\bibitem[{{Navarro} {et~al.}(1996){Navarro}, {Frenk}, \& {White}}]{NFW}
{Navarro}, J.~F., {Frenk}, C.~S., \& {White}, S. D.~M. 1996, ApJ, 462, 563

\bibitem[{{Peccei} \& {Quinn}(1977)}]{Peccei1977}
{Peccei}, R.~D., \& {Quinn}, H.~R. 1977, \prl, 38, 1440

\bibitem[{Preskill {et~al.}(1983)Preskill, Wise, \& Wilczek}]{preskill1983}
Preskill, J., Wise, M.~B., \& Wilczek, F. 1983, Phys.\ Lett.\ B, 120, 127

\bibitem[{{Price-Whelan} {et~al.}(2018){Price-Whelan}, {Sip{\H{o}}cz},
  {G{\"u}nther}, {Lim}, {Crawford}, {Conseil}, {Shupe}, {Craig}, {Dencheva},
  {Ginsburg}, {VanderPlas}, {Bradley}, {P{\'e}rez-Su{\'a}rez}, {de Val-Borro},
  {Paper Contributors}, {Aldcroft}, {Cruz}, {Robitaille}, {Tollerud},
  {Coordination Committee}, {Ardelean}, {Babej}, {Bach}, {Bachetti}, {Bakanov},
  {Bamford}, {Barentsen}, {Barmby}, {Baumbach}, {Berry}, {Biscani}, {Boquien},
  {Bostroem}, {Bouma}, {Brammer}, {Bray}, {Breytenbach}, {Buddelmeijer},
  {Burke}, {Calderone}, {Cano Rodr{\'\i}guez}, {Cara}, {Cardoso}, {Cheedella},
  {Copin}, {Corrales}, {Crichton}, {D{\textquoteright}Avella}, {Deil},
  {Depagne}, {Dietrich}, {Donath}, {Droettboom}, {Earl}, {Erben}, {Fabbro},
  {Ferreira}, {Finethy}, {Fox}, {Garrison}, {Gibbons}, {Goldstein}, {Gommers},
  {Greco}, {Greenfield}, {Groener}, {Grollier}, {Hagen}, {Hirst}, {Homeier},
  {Horton}, {Hosseinzadeh}, {Hu}, {Hunkeler}, {Ivezi{\'c}}, {Jain}, {Jenness},
  {Kanarek}, {Kendrew}, {Kern}, {Kerzendorf}, {Khvalko}, {King}, {Kirkby},
  {Kulkarni}, {Kumar}, {Lee}, {Lenz}, {Littlefair}, {Ma}, {Macleod},
  {Mastropietro}, {McCully}, {Montagnac}, {Morris}, {Mueller}, {Mumford},
  {Muna}, {Murphy}, {Nelson}, {Nguyen}, {Ninan}, {N{\"o}the}, {Ogaz}, {Oh},
  {Parejko}, {Parley}, {Pascual}, {Patil}, {Patil}, {Plunkett}, {Prochaska},
  {Rastogi}, {Reddy Janga}, {Sabater}, {Sakurikar}, {Seifert}, {Sherbert},
  {Sherwood-Taylor}, {Shih}, {Sick}, {Silbiger}, {Singanamalla}, {Singer},
  {Sladen}, {Sooley}, {Sornarajah}, {Streicher}, {Teuben}, {Thomas},
  {Tremblay}, {Turner}, {Terr{\'o}n}, {van Kerkwijk}, {de la Vega}, {Watkins},
  {Weaver}, {Whitmore}, {Woillez}, {Zabalza}, \& {Contributors}}]{astropy:2018}
{Price-Whelan}, A.~M., {Sip{\H{o}}cz}, B.~M., {G{\"u}nther}, H.~M., {et~al.}
  2018, AJ, 156, 123

\bibitem[{Safdi {et~al.}(2019)Safdi, Sun, \& Chen}]{Safdi2019}
Safdi, B.~R., Sun, Z., \& Chen, A.~Y. 2019, Physical Review D, 99,
  arXiv:1811.01020

\bibitem[{{Shifman} {et~al.}(1980){Shifman}, {Vainshtein}, \&
  {Zakharov}}]{Shifman1980}
{Shifman}, M.~A., {Vainshtein}, A.~I., \& {Zakharov}, V.~I. 1980, Nuclear
  Physics B, 166, 493

\bibitem[{{Sikivie}(1983)}]{sikivie1983}
{Sikivie}, P. 1983, \prl, 51, 1415

\bibitem[{{van der Walt} {et~al.}(2011){van der Walt}, {Colbert}, \&
  {Varoquaux}}]{NumPy}
{van der Walt}, S., {Colbert}, S.~C., \& {Varoquaux}, G. 2011, Computing in
  Science Engineering, 13, 22

\bibitem[{Weinberg(1978)}]{weinberg1978}
Weinberg, S. 1978, Phys.\ Rev.\ Lett., 40, 223

\bibitem[{Wilczek(1978)}]{wilczek1978}
Wilczek, F. 1978, Phys.\ Rev.\ Lett., 40, 279

\bibitem[{Zhitnitskij(1980)}]{Zhitnitsky1980}
Zhitnitskij, A.~R. 1980, Yadernaya Fizika, 31, 497

\bibitem[{Zhong {et~al.}(2018)Zhong, {Al Kenany}, Backes, Brubaker, Cahn,
  Carosi, Gurevich, Kindel, Lamoreaux, Lehnert, Lewis, Malnou, Maruyama,
  Palken, Rapidis, Root, Simanovskaia, Shokair, Speller, Urdinaran, \& {Van
  Bibber}}]{Zhong2018}
Zhong, L., {Al Kenany}, S., Backes, K.~M., {et~al.} 2018, Physical Review D,
  97, 092001

\end{thebibliography}

\appendix

%\onecolumngrid

  \section{Spectra of the Magnetar PSR J1745$-$2900\label{sec:appendix}}

\begin{figure}[h]
\begin{centering}
  \includegraphics[scale=0.48,trim=0 0 0 0,clip=true]{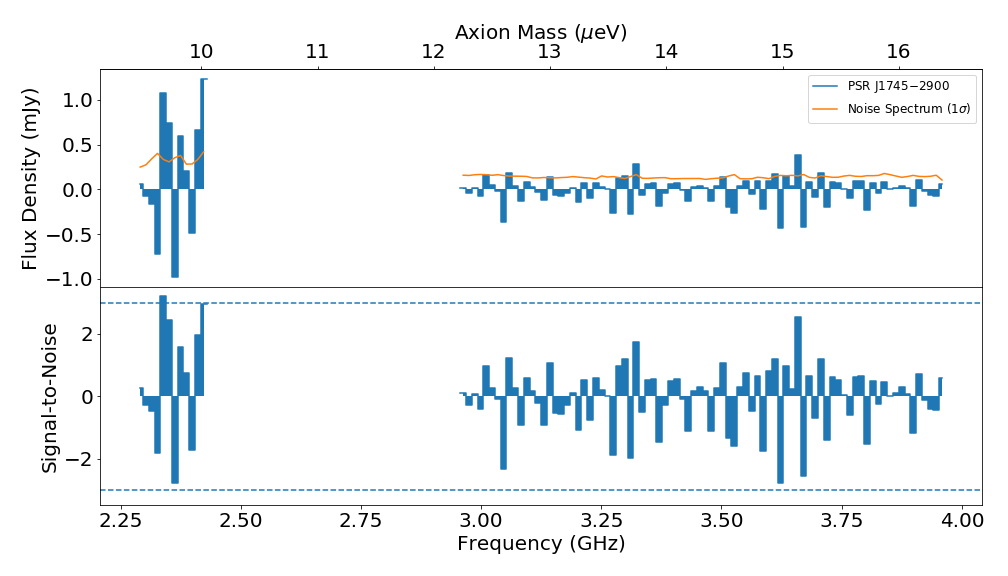}
\caption{%\footnotesize
  S-band flux density, noise, and signal-to-noise spectra.
  The upper spectrum can provide limits on $g_{a\gamma\gamma}$, while the lower spectrum shows the
  significance of spectral features.}  \label{fig:Spec_S}
\end{centering}
\end{figure}

\begin{figure}
\begin{centering}
  \includegraphics[scale=0.48,trim=0 0 0 0,clip=true]{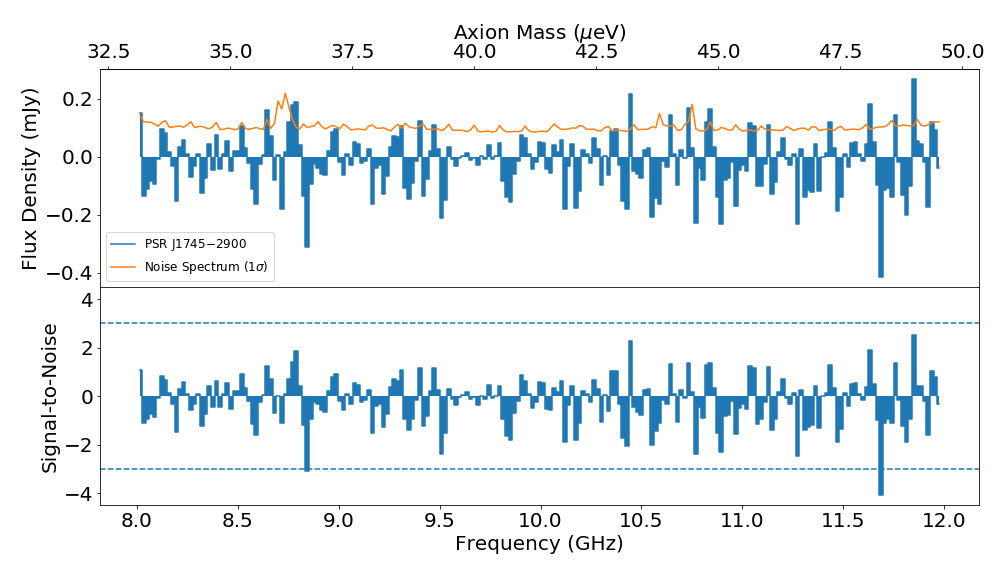}
\caption{%\footnotesize
  X-band flux density, noise, and signal-to-noise spectra from VLA program 15A-418.}  \label{fig:Spec_X}
\end{centering}
\end{figure}

\begin{figure}
\begin{centering}
  \includegraphics[scale=0.48,trim=0 0 0 0,clip=true]{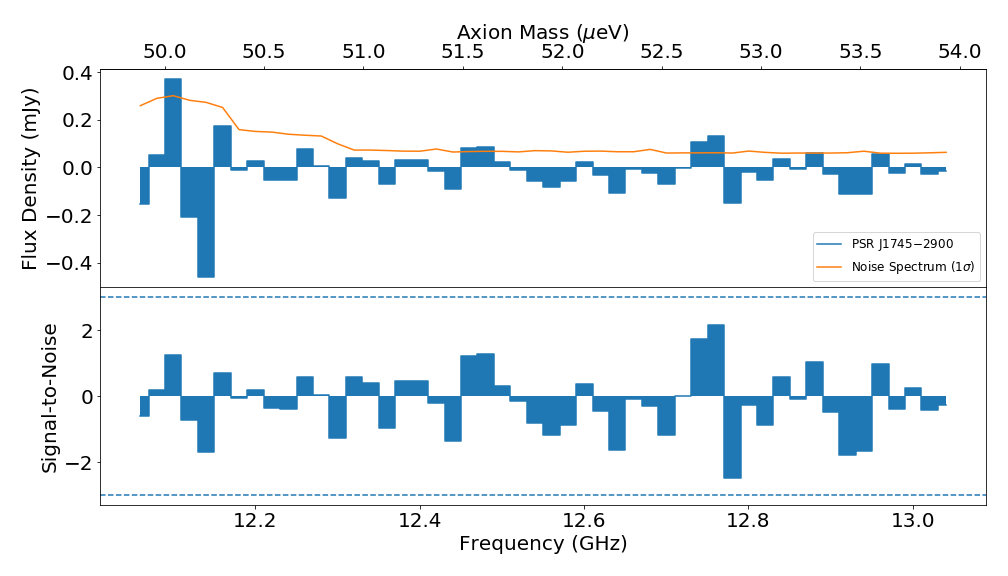}
\caption{%\footnotesize
  Ku-band 12--13~GHz flux density, noise, and signal-to-noise spectra.}  \label{fig:Spec_Ku12}
\end{centering}
\end{figure}

\begin{figure}
\begin{centering}
  \includegraphics[scale=0.48,trim=0 0 0 0,clip=true]{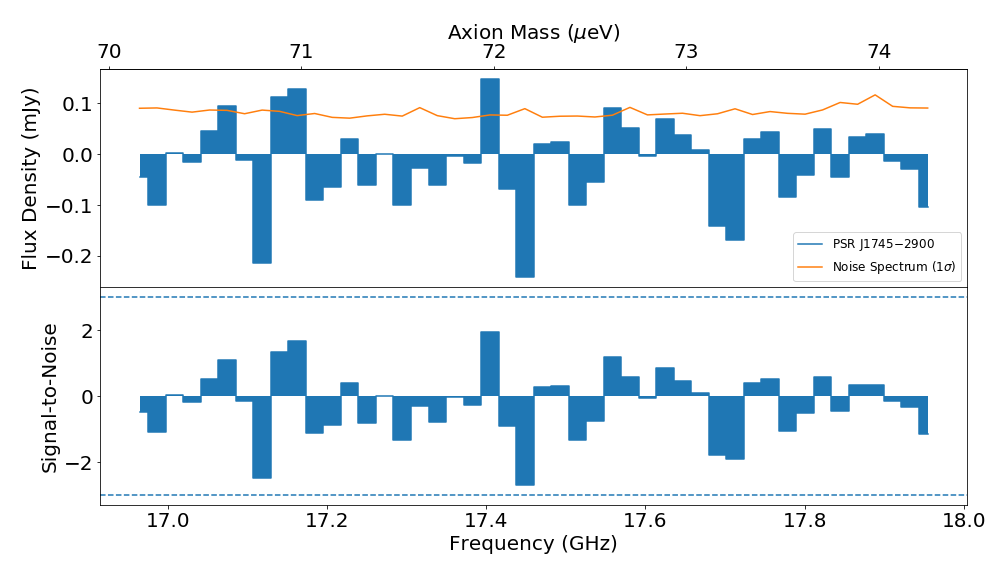}
\caption{%\footnotesize
  Ku-band 17--18~GHz flux density, noise, and signal-to-noise spectra.}  \label{fig:Spec_Ku17}
\end{centering}
\end{figure}

\begin{figure}
\begin{centering}
  \includegraphics[scale=0.48,trim=0 0 0 0,clip=true]{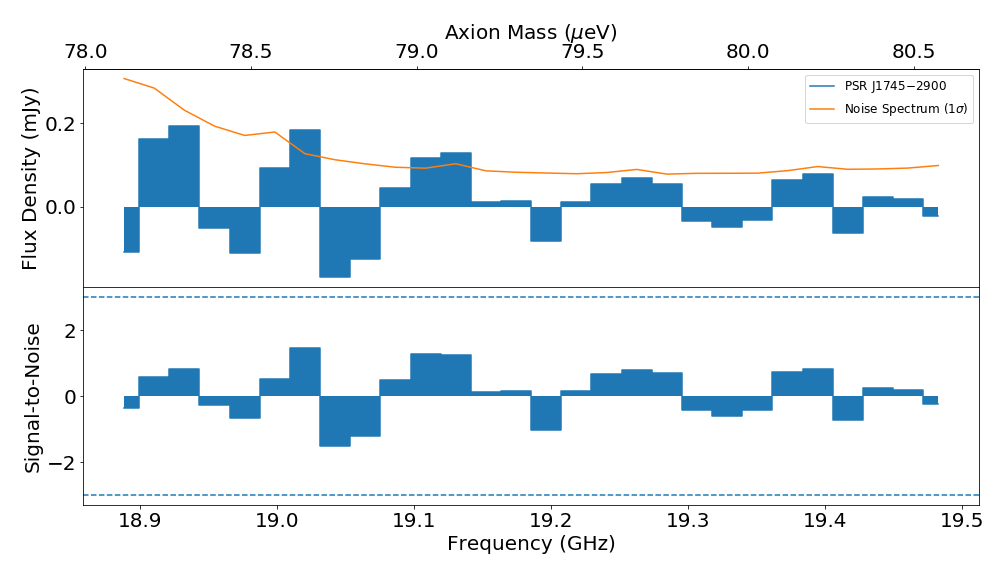}
\caption{%\footnotesize
  K-band 18.9--19.5~GHz flux density, noise, and signal-to-noise spectra.}  \label{fig:Spec_K19}
\end{centering}
\end{figure}

\begin{figure}
\begin{centering}
  \includegraphics[scale=0.48,trim=0 0 0 0,clip=true]{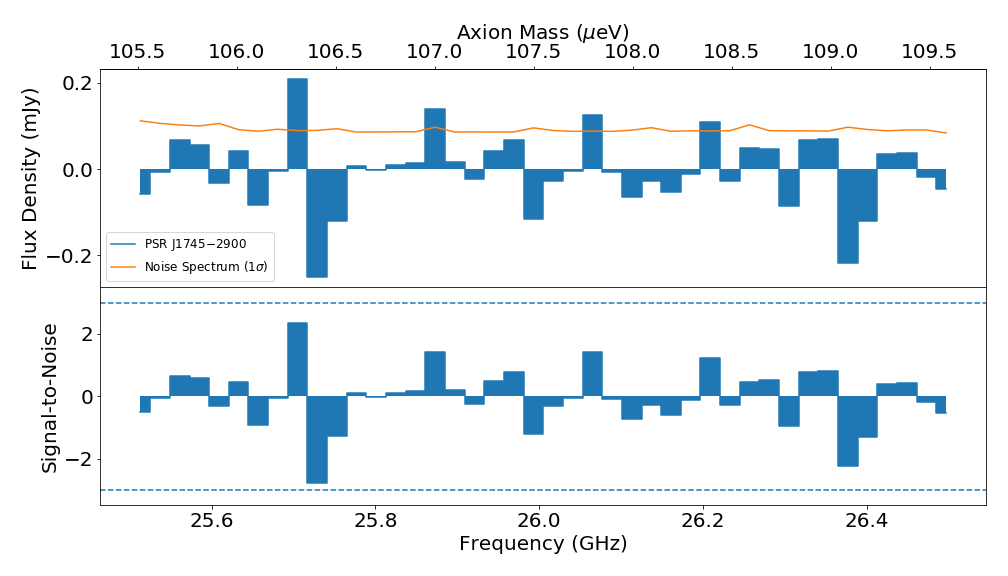}
\caption{%\footnotesize
  K-band 25.5--26.5~GHz flux density, noise, and signal-to-noise spectra.}  \label{fig:Spec_K26}
\end{centering}
\end{figure}

\begin{figure}
\begin{centering}
  \includegraphics[scale=0.48,trim=0 0 0 0,clip=true]{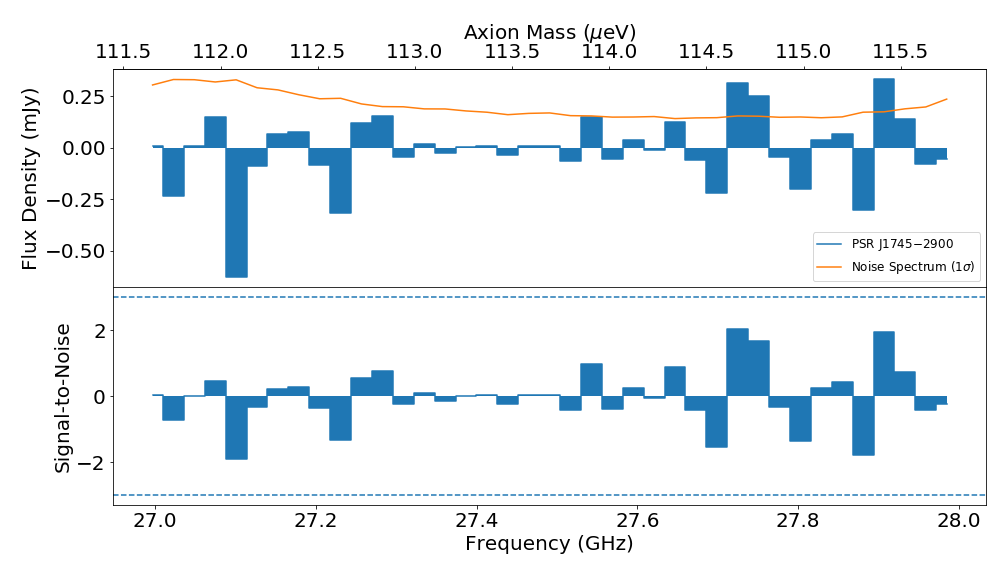}
\caption{%\footnotesize
  Ka-band 27--28~GHz flux density, noise, and signal-to-noise spectra.}  \label{fig:Spec_Ka27}
\end{centering}
\end{figure}

\begin{figure}
\begin{centering}
  \includegraphics[scale=0.48,trim=0 0 0 0,clip=true]{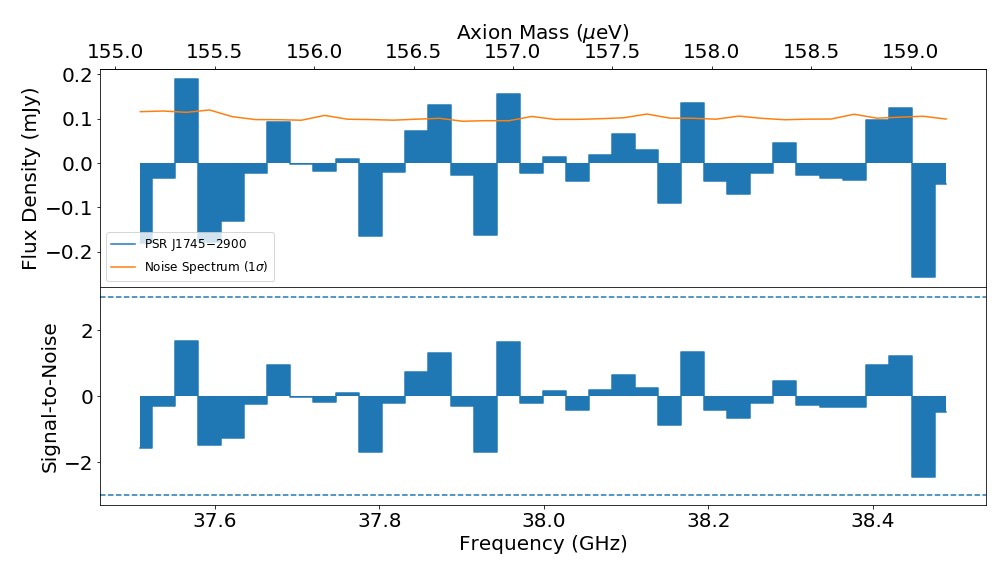}
\caption{%\footnotesize
  Ka-band 37.5--38.5~GHz flux density, noise, and signal-to-noise spectra. }
    \label{fig:Spec_Ka38}
\end{centering}
\end{figure}

\begin{figure}
\begin{centering}
  \includegraphics[scale=0.48,trim=0 0 0 0,clip=true]{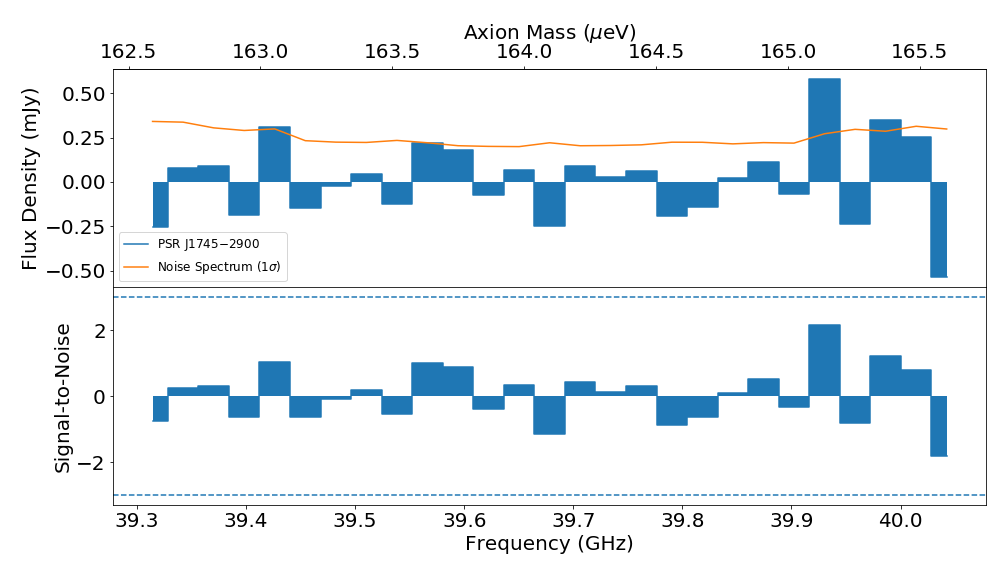}
\caption{%\footnotesize
  Q-band 39.3--40~GHz flux density, noise, and signal-to-noise spectra.}  \label{fig:Spec_Q}
\end{centering}
\end{figure}

  Here we present the new radio spectra of the individual bands 
  used to derive limits on the axion-photon coupling $g_{a\gamma\gamma}$ versus
  axion mass $m_a$ presented in the main Letter.  
  Figures \ref{fig:Spec_S}--\ref{fig:Spec_Q} show the flux density spectra, noise spectra, and
  significance spectra used for flux density and $g_{a\gamma\gamma}$ limits (Figures \ref{fig:flux} and \ref{fig:limits}).
  Spectra obtained from VLA
  programs 14A-231 and 14A-232 (L-, C-, X-, Ku-, and Ka-bands listed in Table \ref{tab:obs}) are presented
  in \citet{darling2020}.

\end{document}